\documentclass[reprint,amsmath,amssymb,aip,eqsecnum,jcp]{revtex4-1}
\usepackage[dvips]{graphicx}
\usepackage{color}
\usepackage{amsmath}
\usepackage{amssymb}
\usepackage{pstricks,pst-grad}
\usepackage{graphicx}
\usepackage{amsbsy}
\usepackage{epic}

\usepackage[linktoc=all,unicode=true,pdfusetitle,  bookmarks=true,bookmarksnumbered=false,bookmarksopen=false, breaklinks=true,pdfborder={0 0 0},backref=false,colorlinks=true]{hyperref}
\hypersetup{linkcolor=blue,citecolor=red}

\usepackage[hyphenbreaks]{breakurl}

\usepackage[normalem]{ulem}

\usepackage{dcolumn}
\usepackage{bm}

\newcommand\beq{\begin{equation}}
\newcommand\eeq{\end{equation}}
\newcommand\beqa{\begin{eqnarray}}
\newcommand\eeqa{\end{eqnarray}}

\newcommand{\nn}{\nonumber\\}
\newcommand\eff{\text{eff}}
\newcommand\dep{\text{dep}}
\newcommand\hs{\text{HS}}
\newcommand\etar{\eta_{p,\text{r}}}
\newcommand{\rr}{\mathbf{r}}

\def\bal#1\eal{\begin{align}#1\end{align}}

\newcommand{\xx}{5}
\newcommand{\yy}{5}

\newcommand{\xmax}{25}
\newcommand{\ymax}{22}

\newcommand{\dE}{\begin{picture}(\xmax,\ymax)(-\xx,\yy)
\thicklines
\setlength{\unitlength}{.1mm}
\put(0,60){\circle*{18}}
\put(60,60){\circle*{18}}
\put(0,0){\circle*{18}}
\put(60,0){\circle*{18}}
\put(9,60){\line(1,0){42}}
\put(9,0){\line(1,0){42}}
\put(0,9){\line(0,1){42}}
\put(60,9){\line(0,1){42}}
\end{picture}}

\newcommand{\DE}{\begin{picture}(\xmax,\ymax)(-\xx,\yy)
\setlength{\unitlength}{.1mm}
\put(0,60){\circle*{18}}
\put(60,60){\circle*{18}}
\put(0,0){\circle*{18}}
\put(60,0){\circle*{18}}
\put(9,60){\line(1,0){42}}
\dottedline{5}(9,0)(51,0)
\put(0,9){\line(0,1){42}}
\put(60,9){\line(0,1){42}}
\end{picture}}

\newcommand{\DDE}{\begin{picture}(\xmax,\ymax)(-\xx,\yy)
\setlength{\unitlength}{.1mm}
\put(0,60){\circle*{18}}
\put(60,60){\circle*{18}}
\put(0,0){\circle{18}}
\put(60,0){\circle{18}}
\put(9,60){\line(1,0){42}}
\put(0,9){\line(0,1){42}}
\put(60,9){\line(0,1){42}}
\end{picture}}

\newcommand{\dG}{\begin{picture}(\xmax,\ymax)(-\xx,\yy)
\thicklines
\setlength{\unitlength}{.1mm}
\put(0,60){\circle*{18}}
\put(60,60){\circle*{18}}
\put(0,0){\circle*{18}}
\put(60,0){\circle*{18}}
\put(9,60){\line(1,0){42}}
\put(9,0){\line(1,0){42}}
\put(0,9){\line(0,1){42}}
\put(60,9){\line(0,1){42}}
\put(7,7){\line(1,1) {46.5}}
\end{picture}}

\newcommand{\DG}{\begin{picture}(\xmax,\ymax)(-\xx,\yy)
\setlength{\unitlength}{.1mm}
\put(0,60){\circle*{18}}
\put(60,60){\circle*{18}}
\put(0,0){\circle*{18}}
\put(60,0){\circle*{18}}
\put(9,60){\line(1,0){42}}
\dottedline{5}(9,0)(51,0)
\put(0,9){\line(0,1){42}}
\put(60,9){\line(0,1){42}}
\put(7,7){\line(1,1) {46.5}}
\end{picture}}

\newcommand{\DDG}{\begin{picture}(\xmax,\ymax)(-\xx,\yy)
\setlength{\unitlength}{.1mm}
\put(0,60){\circle*{18}}
\put(60,60){\circle*{18}}
\put(0,0){\circle{18}}
\put(60,0){\circle{18}}
\put(9,60){\line(1,0){42}}
\put(0,9){\line(0,1){42}}
\put(60,9){\line(0,1){42}}
\put(7,7){\line(1,1) {46.5}}
\end{picture}}

\newcommand{\DI}{\begin{picture}(\xmax,\ymax)(-\xx,\yy)
\setlength{\unitlength}{.1mm}
\put(0,60){\circle*{18}}
\put(60,60){\circle*{18}}
\put(0,0){\circle*{18}}
\put(60,0){\circle*{18}}
\dottedline{5}(9,0)(51,0)
\put(0,9){\line(0,1){42}}
\put(60,9){\line(0,1){42}}
\put(7,7){\line(1,1) {46.5}}
\put(7,53){\line(1,-1){46.5}}
\end{picture}}

\newcommand{\DDI}{\begin{picture}(\xmax,\ymax)(-\xx,\yy)
\setlength{\unitlength}{.1mm}
\put(0,60){\circle*{18}}
\put(60,60){\circle*{18}}
\put(0,0){\circle{18}}
\put(60,0){\circle{18}}
\put(0,9){\line(0,1){42}}
\put(60,9){\line(0,1){42}}
\put(7,7){\line(1,1) {46.5}}
\put(7,53){\line(1,-1){46.5}}
\end{picture}}

\newcommand{\dL}{\begin{picture}(\xmax,\ymax)(-\xx,\yy)
\thicklines
\setlength{\unitlength}{.1mm}
\put(0,60){\circle*{18}}
\put(60,60){\circle*{18}}
\put(0,0){\circle*{18}}
\put(60,0){\circle*{18}}
\put(9,60){\line(1,0){42}}
\put(9,0){\line(1,0){42}}
\put(0,9){\line(0,1){42}}
\put(60,9){\line(0,1){42}}
\put(7,7){\line(1,1){46.5}}
\put(7,53){\line(1,-1){46.5}}
\end{picture}}

\newcommand{\DL}{\begin{picture}(\xmax,\ymax)(-\xx,\yy)
\setlength{\unitlength}{.1mm}
\put(0,60){\circle*{18}}
\put(60,60){\circle*{18}}
\put(0,0){\circle*{18}}
\put(60,0){\circle*{18}}
\put(9,60){\line(1,0){42}}
\dottedline{5}(9,0)(51,0)
\put(0,9){\line(0,1){42}}
\put(60,9){\line(0,1){42}}
\put(7,7){\line(1,1){46.5}}
\put(7,53){\line(1,-1){46.5}}
\end{picture}}

\newcommand{\DDL}{\begin{picture}(\xmax,\ymax)(-\xx,\yy)
\setlength{\unitlength}{.1mm}
\put(0,60){\circle*{18}}
\put(60,60){\circle*{18}}
\put(0,0){\circle{18}}
\put(60,0){\circle{18}}
\put(9,60){\line(1,0){42}}
\put(0,9){\line(0,1){42}}
\put(60,9){\line(0,1){42}}
\put(7,7){\line(1,1){46.5}}
\put(7,53){\line(1,-1){46.5}}
\end{picture}}

\begin{document}



\title{The effective colloid interaction in the Asakura--Oosawa model. Assessment of non-pairwise terms from the virial expansion}


\author{Andr\'es Santos}
\email{andres@unex.es}
\homepage{http://www.unex.es/eweb/fisteor/andres/}
\affiliation{Departamento de F\'{\i}sica and Instituto de Computaci\'on Cient\'ifica Avanzada (ICCAEx), Universidad de
Extremadura,  E-06071 Badajoz, Spain}
\author{Mariano L\'{o}pez de Haro}
\email{malopez@unam.mx}
\homepage{http://xml.ier.unam.mx/xml/tc/ft/mlh/}
\affiliation{Instituto de Energ\'{\i}as Renovables, Universidad Nacional Aut\'onoma de M\'exico (U.N.A.M.),
Temixco, Morelos 62580, M{e}xico}
\author{Giacomo Fiumara}
\email{giacomo.fiumara@unime.it}
\affiliation{
Department of Mathematics and Computer Science, University of Messina,
Viale F. Stagno D'Alcontres 31, I-98166 Messina, Italy}
\author{Franz Saija}
\email{franz.saija@cnr.it}
\affiliation{CNR-IPCF, Viale F. Stagno d'Alcontres, 37-98158 Messina, Italy}
\date{\today}

\begin{abstract}
The relevance of neglecting three- and four-body interactions in the coarse-grained version of the Asakura--Oosawa model is examined. A mapping between the first few virial coefficients of the binary nonadditive hard-sphere mixture representative of this model and those arising from the {coarse-grained} (pairwise) depletion potential approximation allows for a quantitative evaluation of the effect of such interactions. This turns out to be  especially important for large size ratios and large reservoir polymer packing fractions.
\end{abstract}

\maketitle

\section{Introduction}

The description of the thermodynamic properties of complex fluids is in general not an easy task. In it, one faces the presence of many degrees of freedom and maybe also of different length and time scales. An approach that is usually followed is to recur to coarse-graining. In this approach, what one attempts is to integrate out the irrelevant degrees of freedom and hence to end up with a simpler (equivalent) system with an effective interaction that hopefully captures exactly the essential features of the real interaction. Achieving an exact coarse-graining is, however, also difficult. This is due to the fact that, even if the underlying original molecular interactions are pairwise, the resulting effective potential turns out to be in general a many-body one. For this reason, in the coarse-graining process it is usual to replace the full many-body potential by a simpler effective one in which only pair interactions are involved. The question then arises as to whether the thermodynamic properties derived with the effective potential provide a reliable account of the same properties for the original fluid.

In the early {20th} century,  {Kamerlingh Onnes}\cite{KO01}  introduced the (then empirical) virial series to provide a mathematical representation of experimental pressure-density-temperature data of gases and liquids. In a broader context, one refers to a virial expansion of a given property when such {a} property is expressed as a power series in density. Thus, the virial expansion represents in principle a systematic way for calculating the properties of bulk matter, provided of course that the coefficients in the expansion (the {so-called} virial coefficients) are known accurately either through direct measurement or from theoretical developments. At least formally, one of the great achievements of statistical mechanics, and a major breakthrough in the theoretical approach to calculating virial coefficients involved in the equation of state of fluids, occurred when Mayer\cite{M37,MGM40} was able to obtain general expressions for the corresponding virial coefficients in terms of sums of cluster integrals over the interaction among groups of fluid particles. {In particular,} the second virial coefficient depends only on pair interactions, while the third virial coefficient depends on two- and three-body interactions, and so on. If the intermolecular potential is simple enough, some  of such virial coefficients may be calculated analytically. For instance, in the case of the hard-sphere {(HS)} fluid, the first four virial coefficients are known analytically. The same applies to the celebrated Asakura--Oosawa (AO) model,\cite{AO54,AO58,V76} which describes colloidal {HSs} in a solvent of ideal polymers that interpenetrate each other but interact with the colloids via a {HS} repulsion. In this case, considering that the system may be taken as a nonadditive hard-sphere {(NAHS) binary} mixture with high size asymmetry, analytical results for the first four virial coefficients have been very recently reported.\cite{HTSYFS15} In general, however, numerical evaluation is required and even for simple potentials such as the ones just mentioned there are various technical difficulties involved in computing the higher order virial coefficients. A noteworthy aspect of the {usefulness} of virial coefficients is that the comparison of these coefficients for the original system and the ones stemming out of the coarse-grained interaction potential will indicate to what extent and under which conditions the neglect of the many-body terms in the latter has an impact on whether the thermodynamic properties of both systems agree.

{Although largely ignored for about 20 years, interest in the AO model grew in the 1970s and 1980s and it started to get significant attention in the 1990s, which continues up to this day}.\cite{GHR83,MF91,LPPSW92,MF94,IOPP95,DBE99,DRE99a,DRE00,L01c,BES03,TRK03,VVPR06,FT08,RE08,AWRE11,LT11,BVS14,TF14,AW14a,AW14b,HTSYFS15,FGS15} It is well known that its coarse-grained description involves an effective (depletion) pairwise interaction between the colloids that, among other things, leads to fluid-fluid demixing.  Attempts to examine whether the thermodynamics obtained with the depletion potential agrees with the one of the full mixture have also been reported. In particular,
{it was found that, for a polymer/colloid size ratio $q$ smaller than the threshold value $q_0=2/\sqrt{3}-1\approx 0.1547$, the AO pair depletion potential turns out to be the only one contributing to the exact effective interaction among the solutes.\cite{DBE99} Recently, Ashton and Wilding\cite{AW14a,AW14b}  focused, via simulation,} on the dominant many-body effect neglected in the pair potential description {if $q>q_0$}, namely the one associated {with} the interaction between three colloidal particles. To this end, they examined the difference between the third virial coefficient of the full system and that of the effective system.

Here we will follow a similar route but also profit from the availability of the first five virial coefficients of the full {NAHS} mixture corresponding to the AO model\cite{HTSYFS15} to derive, by the exact mapping that may be performed between both sets of coefficients, partial contributions to the second, third, and fourth virial coefficients of the effective one-component colloidal fluid. The aim is to quantify the deviations from the exact results that one gets for both the third and fourth virial coefficients when computing them with the coarse-grained depletion potential. {As we will see, the influence of non-pairwise interactions on the third and fourth virial coefficients is rather small for $q\lesssim 0.4$ but becomes increasingly important for larger values of the size ratio.}

The paper is organized as follows. In {Sec.\ \ref{sec2}}, and in order to make the paper self-contained, we recall the results for the first five virial coefficients of the {original} AO {binary} mixture. This is followed in Sec.\ \ref{sec3} by the introduction of the osmotic pressure of the colloidal system, which allows us to make the mapping between the virial coefficients of the mixture and those coming out of the {effective one-component colloidal system}. The analytical results that follow from the {(coarse-grained)} {pair depletion potential approximation} are presented in Sec.\ \ref{sec4}. {Section \ref{sec5} provides a comparison between the exact and the approximate results.} The paper is closed in Sec.\ {\ref{sec6}} with further discussion and some concluding remarks.

\section{Virial coefficients of the AO model}
\label{sec2}

Consider a binary fluid mixture of {$N=N_c+ N_p$} {spheres} (colloids+polymers)  {in a}  volume $V$. {The colloid and polymer mole fractions are $x_c=N_c/N$ and $x_p=N_p/N=1-x_c$, respectively. Analogously, the partial and total number densities are $\rho_c=N_c/V$, $\rho_p=N_p/V$, and $\rho=\rho_c+\rho_p=N/V$. The interactions are assumed to be of HS type. The} distance of closest approach between spheres of species {$\alpha$ and $\gamma$}, denoted by {$\sigma_{\alpha\gamma}$}, is such that {$\sigma_{cc}=\sigma_c$, $\sigma_{pp}=0$, and $\sigma_{cp}=\frac{1}{2}\sigma_c(1+q)$}, with the size ratio $q$  acting as the (positive) nonadditivity parameter.
{The colloid packing fraction is $\eta_c=\frac{\pi}{6}\rho_c\sigma_c^3$}. {For simplicity, from now on we choose $\sigma_c=1$ as the unit of length.} This NAHS mixture  {defines} the well known AO model.\cite{BVS14}

The usual virial expansion  {of the mixture} reads
\bal
{\beta a(\rho_c,\rho_p)}=&{\rho_c\ln\left(\rho_c\Lambda_c^3\right)+\rho_p\ln\left(\rho_p\Lambda_p^3\right)-\rho}\nn
&{+\sum_{n=2}^{\infty}\frac{B_n({x_c},q)}{n-1} \rho^{n}},
\label{free_e}
\eal
\beq
{\beta p(\rho_c,\rho_p)=\rho+\sum_{n=2}^{\infty}B_n({x_c},q) \rho^{n}},
\label{comp}
\eeq
where {$a$ is the free energy per unit volume,} $p$ is the pressure, {$\beta=1/k_BT$ ($k_B$ being the Boltzmann constant and $T$ being the absolute temperature)}, {$\Lambda_\alpha$ is the thermal de Broglie wavelength of species $\alpha$,} and {the notation $B_n(x_c,q)$ makes it} explicit that the virial coefficients  depend only on {the mole fraction {$x_c$} of the colloids} and on the size ratio $q$.

The second, third, fourth, and fifth virial coefficients of the AO model are given by\cite{HTSYFS15}
\beq
B_2{({x_c},q)}={x_c^2} B_{11}+2{x_c}{x_p}B_{12}{(q)},
\label{B2}
\eeq
\beq
B_3{({x_c},q)}={x_c^3}C_{111}+3{x_c^2}{x_p}C_{112}{(q)},
\label{B3}
\eeq
\beq
B_4{({x_c},q)}={x_c^4}D_{1111}+4{x_c^3}{x_p}D_{1112}{(q)}+6{x_c^2}{x_p^2}D_{1122}{(q)},
\label{B4}
\eeq
\bal
B_5{({x_c},q)}=&{x_c^5}E_{11111}+5{x_c^4}{x_p}E_{11112}{(q)}\nn
&+10{x_c^3}{x_p^2} E_{11122}{(q)}
{+10{x_c^2}{x_p^3} E_{11222}{(q)}},
\label{B5}
\eal
where {all} {the composition-independent coefficients}, {except $E_{11112}$ and $E_{11122}$}, are exactly known {as functions of $q$},
\beq
B_{11}=\frac{\pi}{6}{4},\quad
\label{B11}
B_{12}=\frac{\pi}{6}\frac{(1+q)^3}{2},
\eeq
\beq
C_{111}= \left(\frac{\pi}{6}\right)^2{10},\quad
C_{112}=\left(\frac{\pi}{6}\right)^2\frac{1+6q+15q^2+8q^3}{3},
\label{C112}
\eeq
\beq
D_{1111}=\left(\frac{\pi}{6}\right)^3 \left(\frac{2707}{70}+\frac{219\sqrt{2}}{35\pi}-\frac{4131\cos^{-1}\frac{1}{3}}{70\pi}\right),
\label{D1111}
\eeq
\beq
D_{1122}=-\left(\frac{\pi}{6}\right)^3{q^5}\left(\frac{27}{20}+\frac{12q}{5}
+\frac{51q^2}{35}+\frac{51q^3}{140}+\frac{17q^4}{420}\right),
\label{D1122}
\eeq
\beq
D_{1112}=\begin{cases}
D_{1112}^{{(a)}},&q\leq q_0,\\
D_{1112}^{{(a)}}+D_{1112}^{{(b)}},&q>q_0,
\end{cases}
\eeq
\bal
D_{1112}^{{(a)}}=&\left(\frac{\pi}{6}\right)^3\left(\frac{1}{4}+\frac{9q}{4}+9q^2
+\frac{21q^3}{4}+\frac{27q^4}{8}+\frac{27q^5}{40}\right.
\nn
&
\left.-\frac{27q^6}{5}-\frac{162q^7}{35}-\frac{81q^8}{56}-\frac{9q^9}{56}\right),
\label{6a}
\eal
\bal
D_{1112}^{{(b)}}=&\left(\frac{\pi}{6}\right)^3\frac{1}{280\pi}\Big[\frac{Q}{12}\left(10Q^6-
51Q^4+210Q^2+6976\right)
\nn
&
-486P_1(Q^2+9)+\frac{q+1}{3} P_2\left(5Q^8-28Q^6\right.\nn
& \left.+129Q^4-124Q^2
+11378\right)
\Big],
\label{DeltaB4}
\eal
\beq
E_{11111}=\left(\frac{\pi}{6}\right)^4b_5,\quad b_5\simeq 28.224512,
\eeq
\bal
{E_{11222}=}&{-\left(\frac{\pi}{6}\right)^4 \frac{q^7}{8400}\left(3240+7695q+6780q^2+2706q^3\right.}\nn
&{\left.+492q^4+41q^5\right)}.
\label{E11222}
\eal
In Eq.\ \eqref{DeltaB4}, $Q\equiv\sqrt{3q^2+6q-1}$, $P_1\equiv \tan^{-1} Q$, and $P_2\equiv \tan^{-1}\left[Q/(q+1)\right]$.

To our knowledge there are no analytical results for the composition-independent coefficients $E_{11112}$ and $E_{11122}$ for general values of $q$.
Therefore, we have  computed them by {a} standard Monte Carlo (MC) numerical integration procedure for a number of values of $q$ in the range $0.05\leq q\leq 1$. The results are displayed in Table \ref{tab1}, {which is more extensive than the equivalent table of Ref.\ \onlinecite{HTSYFS15}.}

{There exist approximate analytical theories, like the free volume (FV) theory,\cite{LPPSW92} that account in closed form for the equation of state of the full AO mixture. For further use, Appendix \ref{appFV} provides the approximate expressions of the first few virial coefficients arising from the FV theory.}

\begin{table}
\caption{Numerical values of the partial coefficients $E_{11112}$  and $E_{11122}$  for some values of the size ratio $q$. The error on the
last significant figure is enclosed in parentheses.
\label{tab1}}
\begin{ruledtabular}
\begin{tabular}{ccc}
 $q$ &  $E_{11112}$ & $E_{11122}$ \\
\hline
 $0.05$ &  $0.0267(6)$  &  $\approx -5 \times 10^{-8}$ \\
 $0.10$ &   $0.0437(4)$   &  $-4.3(9)\times 10^{-6}$              \\
 $0.15$ &   $0.0666(8)$   &$-3.4(8)\times 10^{-5}$               \\
  $0.20$ &   $0.0955(9)$  &     $-1.6(2)\times 10^{4}$                \\
 $0.30$  &  $0.177(1)$       &$-0.00147(4)$                \\
 $0.40$ &   $0.296(4)$     &$-0.0075(2)$                 \\
 $0.50$ &   $0.457(2)$    &$-0.0276(5)$                \\
$0.56$ &   $0.575(5)$       &$-0.0546(5)$              \\
 $0.60$  &  $0.666(3)$     &  $-0.0831(9)$              \\
 $0.70$  &  $0.931(4)$      & $-0.216(2)$              \\
  $0.80$ &  $1.257(7)$       &$-0.505(4)$           \\
 $0.90$ &  $1.652(6)$       &$-1.085(4)$           \\
  $1.00$&$E_{11111}$&$-2.18(3)$ \\
\end{tabular}
\end{ruledtabular}
\end{table}

\section{Osmotic pressure and virial coefficients of the effective colloidal system}
\label{sec3}
{Equations \eqref{free_e} and \eqref{comp} are  expressed in the canonical ensemble $(N_p,N_c,V,T)$. On the other hand, in order to analyze the effective one-component colloidal fluid, it turns out to be convenient to consider the semi-grand-canonical ensemble $(\mu_p,N_c,V,T)$, where $\mu_p$ is the chemical potential of the polymer component. In that ensemble, the pressure of the mixture can be written as\cite{DBE99}}
 \beq
   {\beta} p(\rho_c,z_p)=z_p+ {\beta} \Pi(\rho_c,z_p),
  \label{1}
  \eeq
where
\beq
{z_p=\frac{e^{\beta \mu_p}}{\Lambda_p^3}}
\label{zp}
\eeq
{is the polymer fugacity and } $\Pi(\rho_c,z_p)$ is the osmotic pressure that takes into account the (formally) \emph{exact} effective colloid-colloid interactions mediated by the polymers.  Its virial expansion is
\beq
{\beta\Pi(\rho_c,z_p)=\rho_c+\sum_{n=2}^\infty B_n^\eff(z_p,q)\rho_c^{n}},
\label{osmotic}
\eeq
{where $B_n^\eff(z_p,q)$ are the  virial coefficients of the effective one-component colloidal fluid.}
The fugacity $z_p$ of the polymer component can be equivalently  represented by the \emph{reservoir} polymer packing fraction $\etar=z_p\frac{\pi}{6}q^3$. {Thus, henceforth we make the change $B_n^\eff(z_p,q)\to B_n^\eff(\etar,q)$.}
The effective virial coefficients $B_n^\eff(\etar,{q})$ can be further expressed as {a series in powers of $\etar$,}
\beq
B_n^\eff(\etar,q)=\sum_{j=0}^\infty B_n^{(j)}{(q)}\etar^j.
  \label{2}
\eeq

Our aim {in this section} is to {provide} the \emph{exact} relations between the effective {one-component} virial coefficients $B_n^{(j)}$ (with $n+j\leq 5$) and the binary-mixture
virial coefficients {of} Eqs.\ \eqref{B2}--\eqref{B5}.
{The details  are given in Appendix \ref{appA} with the results}
\begin{subequations}
\label{6}
 \bal
B_2^{(0)}=&B_{11}, \quad {B_3^{(0)}=C_{111},}\\
{B_4^{(0)}}=&{D_{1111},\quad B_5^{(0)}=E_{11111},}
\eal
\end{subequations}
\beq
B_2^{(1)}=\frac{6}{\pi q^3}\left(\frac{3}{2}C_{112}-2B_{12}^2\right),
\label{7}
\eeq
\beq
B_2^{(2)}=2\left(\frac{6}{\pi q^3}\right)^2D_{1122}, \quad {B_2^{(3)}=\frac{5}{2}\left(\frac{6}{\pi q^3}\right)^3E_{11222},}
\label{8}
\eeq
\beq
B_3^{(1)}=\frac{6}{\pi q^3}\left(\frac{8}{3}D_{1112}-6B_{12}C_{112}+\frac{8}{3}B_{12}^3\right),
\label{11}
\eeq
\beq
B_3^{(2)}=\left(\frac{6}{\pi q^3}\right)^2\left(5 E_{11122}-16B_{12}D_{1122}\right),
\label{12}
\eeq
\bal
B_4^{(1)}=&\frac{6}{\pi q^3}\Big(\frac{15}{4}E_{11112}-8B_{12}D_{1112}-\frac{27}{8}C_{112}^2\nn
&+9B_{12}^2C_{112}-2B_{12}^4\Big),
\label{14}
\eal
{Equations \eqref{6}--\eqref{14} provide the sought relationships between the effective and binary-mixture virial coefficients that account for all the three- and four-body interactions up to the fourth virial coefficients of the effective system.}
{Making use of Eqs.\ \eqref{B11}--\eqref{E11222} and the numerical values of Table \ref{tab1}, one can then know the exact $q$-dependence of the  coefficients $B_n^{(j)}$. Furthermore, in the case of the approximate FV theory, the results are explicitly given by Eqs.\ \eqref{B31FV}--\eqref{B22FV}.}

\section{Virial coefficients arising from the use of the effective pair AO potential}
\label{sec4}
{By integrating out the polymer degrees of freedom, it is possible to derive the formally exact effective many-body interaction potential of the colloids, $\Phi_\eff(\rr^{N_c})\equiv\Phi_\eff(\rr_1,\rr_2,\ldots,\rr_{N_c})$, in the AO model. The result is\cite{DBE99}}
\bal
{\Phi_\eff(\rr^{N_c})}=&{-z_p\int d\rr\,\prod_{i=1}^{N_c}\left[1-\Theta\left(\frac{1+q}{2}-|\rr-\rr_i|\right)\right]}\nn
&{+\sum_{i=1}^{N_c-1}\sum_{j=i+1}^{N_c}\phi_\hs(r_{ij})},
\eal
{where $\Theta(x)$ is the Heaviside step function and $\phi_\hs(r)$ is the original colloid-colloid HS pair potential of diameter $\sigma_c=1$. If $q<q_0$, a polymer particle cannot overlap with more than two nonoverlapping colloids, so that $\Phi_\eff$ is exactly given by\cite{DBE99}}
\bal
{\beta\Phi_\eff(\rr^{N_c})=}&{-z_pV\left[1-\eta_c(1+q)^3\right]}\nn
&{+ \sum_{i=1}^{N_c-1}\sum_{j=i+1}^{N_c}\beta\phi_\dep(r_{ij}),}
\label{Phi}
\eal
{where $\phi_\dep(r)$ is the effective  AO pair \emph{depletion} potential. It is given by}
\beq
\beta \phi_{\dep}(r)=
\begin{cases}
\infty,&r<1,\\
-\etar \omega(r),&1<r<1+q,\\
0,&r>1+q,
\end{cases}
\label{16}
\eeq
where
\beq
\omega(r)=\frac{1}{2q^3}(1+q-r)^2(2+2q+r).
\label{17}
\eeq
The corresponding Mayer function is
\bal
f_{\dep}(r)=&f_\hs(r)+\left[e^{\etar\omega(r)}-1\right]\Theta(r-1)\Theta(1+q-r)\nn
=&{f_\hs(r)+\sum_{j=1}^\infty f_\dep^{(j)}(r)\etar^j},
\label{18}
\eal
{with $f_\hs(r)=-\Theta(1-r)$ and $f_\dep^{(j)}(r)=\Theta(r-1)\Theta(1+q-r)[\omega(r)]^j/j!$.

On the other hand, if  $q_0<q\leq 1$, $m$-body terms with $3\leq m\leq 11$ gradually contribute to $\Phi_\eff$, the upper limit ($m=11$) being due to the fact that a polymer can overlap simultaneously with 12 nonoverlapping colloids only if $q>1$.\cite{FGS15} Therefore, Eq.\ \eqref{Phi} becomes an approximation (henceforth referred to as the {\emph{coarse-grained}} depletion approximation) if $q>q_0$. In the remainder of this section we explicitly evaluate the effective virial coefficients $B_{n}^{(j)}$ with $n+j\leq 5$ for any $0\leq q\leq 1$ in this {coarse-grained} approximation.}
\subsection{Second virial coefficient}
As a consequence of \eqref{18}, the second virial coefficient is
\bal
B_2^\eff=&-2\pi\int_0^\infty dr\,r^2f_{{\dep}}(r)\nn
=&\frac{2\pi}{3}-2 \pi\sum_{j=1}^\infty\frac{\etar^j}{j!}\int_1^{1+q}dr\,r^2\left[\omega(r)\right]^j.
\label{19}
\eal
{}From here one can {easily} obtain
\beq
\frac{B_2^{(1)}}{B_2^\hs}=-\frac{3}{2}\left(1+\frac{5q}{4}+\frac{q^2}{2}+\frac{q^3}{12}\right),
\label{20}
\eeq
\beq
\frac{B_2^{(2)}}{B_2^\hs}=-\frac{27}{40q}\left(1+\frac{16q}{9}+\frac{68q^2}{63}+\frac{17q^3}{63}+\frac{17q^4}{567}\right),
\label{21}
\eeq
\bal
\frac{B_2^{(3)}}{B_2^\hs}=&{-\frac{27\pi}{112q^2}\left(1+\frac{19q}{8}+\frac{113q^2}{54}+\frac{451q^3}{540}+\frac{41q^4}{270}\right.}\nn
&{\left.+\frac{41q^5}{3240}\right)},
\label{22}
\eal
{where $B_2^\hs=\frac{2\pi}{3}$}.
Equations \eqref{20}--\eqref{22} agree with Eqs.\ \eqref{7} and \eqref{8}. {Of course, this is an expected result since the pair approximation is exact at the level of the second virial coefficient.}

\subsection{Third virial coefficient}
We now turn to the third virial coefficient
\beq
B_3^\eff=-\frac{(2\pi)^{-3}}{3}\int d\mathbf{k}\,\left[\widetilde{f}_{{\dep}}(k)\right]^3,
\label{23}
\eeq
where
\bal
\widetilde{f}_{{\dep}}(k)=&\int d\mathbf{r}\,e^{-i\mathbf{k}\cdot\mathbf{r}}f_{{\dep}}(r)\nn
=&\frac{4\pi}{k}\int_0^\infty dr\, r\sin(kr) f_{{\dep}}(r)
\label{24}
\eal
is the Fourier transform of $f_{{\dep}}(r)$. {}From Eq.\ \eqref{18} we have
\beq
{\widetilde{f}_{{\dep}}(k)=\widetilde{f}_\hs(k)+\sum_{j=1}^\infty\widetilde{f}_{{\dep}}^{(j)}(k)\etar^j},
\label{25}
\eeq
where
\beq
\widetilde{f}_\hs(k)=\frac{4\pi}{k^3}\left(k\cos k-\sin k\right),
\label{26}
\eeq
\beq
{\widetilde{f}_{{\dep}}^{(j)}(k)=\frac{1}{j!}\frac{4\pi}{k }\int_1^{1+q} dr\, r\sin(kr) \left[\omega(r)\right]^j},
\label{27}
\eeq

According to Eqs.\ \eqref{23} {and \eqref{25}, the coefficients}  $B_3^{(1)}$ and $B_3^{(2)}$ are given by
\beq
B_3^{(1)}=-(2\pi)^{-3}\int d\mathbf{k}\,\left[\widetilde{f}_\hs(k)\right]^2 \widetilde{f}_{{\dep}}^{(1)}(k),
\label{28}
\eeq
\bal
B_3^{(2)}=&-(2\pi)^{-3}\int d\mathbf{k}\,\widetilde{f}_\hs(k)\Big\{
\widetilde{f}_\hs(k)\widetilde{f}_{{\dep}}^{(2)}(k)\nn
&+\left[\widetilde{f}_{{\dep}}^{(1)}(k)\right]^2\Big\}.
\label{29}
\eal
{The explicit expressions of $\widetilde{f}_{{\dep}}^{(1)}(k)$ and $\widetilde{f}_{{\dep}}^{(1)}(k)$ can be obtained from application of Eqs.\ \eqref{17} and \eqref{27} but, for conciseness, they  will be omitted here. Insertion of those expressions into Eqs.\ \eqref{28} and \eqref{29} yields (for $0\leq q\leq 1$)}
\beq
\frac{B_3^{(1)}}{B_3^\hs}=-3\left(1+\frac{4q}{5}-\frac{4q^2}{25}-\frac{14q^3}{75}+\frac{2q^4}{175}+\frac{q^5}{35}+\frac{q^6}{315}\right),
\label{30}
\eeq
\bal
\frac{B_3^{(2)}}{B_3^\hs}=&-\frac{27}{20 q}\left(1-\frac{107q}{90}-\frac{1529q^2}{315}-\frac{9253 q^3}{2520}-\frac{3889q^4}{5670}\right.
\nn
&\left.+\frac{4663 q^5}{18\,900}+\frac{1049 q^6}{9450}+\frac{1049q^7}{113\,400}\right).
\label{31}
\eal

As a byproduct, since Eq.\ \eqref{31} must be exact for $q<q_0$, Eq.\ \eqref{12} allows one to obtain the \emph{exact} expression of $E_{11122}$ for $q<q_0$,
\bal
E_{11122}=&-\frac{\pi^4 q^5}{800}\left({3}+\frac{79q}{18}+\frac{281q^2}{63}+\frac{3529 q^3}{504}+\frac{2519q^4}{378}\right.
\nn
&\left.+\frac{34\,583 q^5}{11\,340}+\frac{3769 q^6}{5670}+\frac{3769q^7}{68\,040}\right).
\label{32}
\eal
The exact values of $E_{11122}$ corresponding to $q=0.05$, $0.10$, and $0.15$ are $-1.229\,60 \times10^{-7}$, $-4.250\,92\times10^{-6}$, and $-3.500\,57\times 10^{-5}$, respectively. By comparison with the third column of Table \ref{tab1}, we observe that the MC results agree with the exact values within the associated uncertainties.

\subsection{Fourth virial coefficient}
Finally, we consider the effective fourth virial coefficient in the {coarse-grained} pair depletion approximation. It is given by
\beq
B_4^\eff=-\frac{1}{8}\left(3\dE+6\dG+\dL\right),
\eeq
where each thick bond represents a Mayer function \eqref{18}. Expanding in powers of $\etar$, one gets
\beq
B_4^{(1)}=-\frac{3}{4}\left(2\DE+4\DG+\DI+\DL\right).
\label{s:d1122}
\eeq
{Now}, a thin solid line between two circles  represents the HS Mayer function $f_\hs(r)$, while a {dotted} line  represents a term $f_{{\dep}}^{(1)}(r)$.
Interestingly, Eq.\ \eqref{s:d1122} can be written as
\beq
B_4^{(1)}=-6\pi\int_1^{1+q}dr\, r^2\omega(r)y_2^\hs(r),
\label{B41}
\eeq
where
\beq
y_2^\hs(r)=\DDE+2\DDG
+\frac{1}{2}\DDI+\frac{1}{2}\DDL
\label{y2HS}
\eeq
is the HS cavity function to second order in density, which is exactly known.\cite{NvH52,SM07}

After some lengthy algebra  it is possible  to find a fully analytical expression for $B_4^{(1)}$ (see Appendix \ref{appB}).
Again, since that expression of $B_4^{(1)}$ is exact for $q\leq q_0$, use of Eq.\ \eqref{14} allows us to derive an exact analytical form of $E_{11112}$ for $q\leq q_0$ [see Eq.\ \eqref{E11112}]. The values  corresponding to $q=0.05$, $0.10$, and $0.15$ are $0.026\,588\,4$,  $0.043\,507\,4$, and $0.066\,359\,4$, respectively. Comparison with the second column of Table \ref{tab1} shows again an excellent agreement of the MC results with the exact values.

\section{Comparison between the exact and the approximate coefficients $B_3^{(1)}$, $B_3^{(2)}$, and $B_4^{(1)}$}
\label{sec5}

\begin{figure}
\includegraphics[width=8cm]{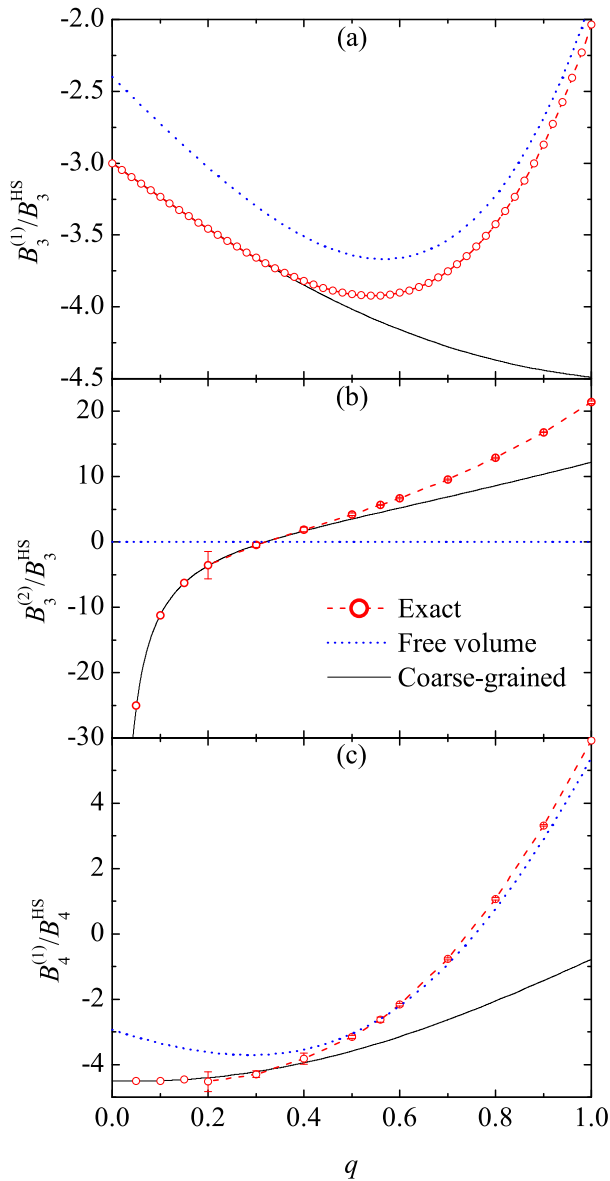}
\caption{Comparison between the exact and the {coarse-grained}  coefficients (a) $B_3^{(1)}$, (b) $B_3^{(2)}$, and (c) $B_4^{(1)}$.}
\label{fig1}
\end{figure}
As said before, {coarse-grained} pair-potential approximation \eqref{Phi} is only correct if $q<q_0\simeq 0.1547$. Beyond that value, the virial coefficients $B_n^\eff$ (with $n\geq 3$) obtained from the AO pair potential \eqref{16} differ from the exact ones. In particular, the exact coefficient $B_3^\eff$ is influenced by three-body interactions,\cite{AW14a,AW14b} while the exact coefficient $B_4^\eff$ is influenced by both three- and four-body interactions.

Here we restrict ourselves to $B_3^{(1)}$, $B_3^{(2)}$, and $B_4^{(1)}$. In the {coarse-grained} approximation, they are given by Eqs.\ \eqref{30}, \eqref{31}, and \eqref{B41depl}--\eqref{DeltaB4b}, respectively (if $q\leq 1$). The exact expressions are given by Eqs.\ \eqref{11}, \eqref{12}, and \eqref{14}, respectively, in terms of the composition-independent virial coefficients of the binary mixture. While in Eq.\ \eqref{11} all the coefficients are known analytically, in Eqs.\ \eqref{12} and \eqref{14} one needs to resort (for $q>q_0$) to numerical MC evaluations  listed in Table \ref{tab1}.
{In the FV theory for the full AO binary mixture, $B_3^{(1)}$ and $B_4^{(1)}$ are given by Eqs.\ \eqref{B31FV} and \eqref{B41FV}, respectively, while $B_3^{(2)}=0$.}

The comparison between the exact and approximate coefficients is carried out in Fig.\ \ref{fig1}. We see that  the influence of three-body interactions on $B_3^{(1)}$ and $B_3^{(2)}$ is practically negligible in the range $q_0\leq q\lesssim 0.4$ but becomes quite important, especially in the case of $B_3^{(1)}$, if $q\gtrsim 0.6$. A similar conclusion can be drawn from $B_4^{(1)}$: the role played by three- and four-body interactions is irrelevant if $q\leq 0.4$ but becomes essential as $q$ increases. We observe that the non-pairwise contributions to the true effective many-body colloid potential tend to increase the values of $B_3^\eff$ and $B_4^\eff$ with respect to the {coarse-grained} estimates, thus partially compensating for the attractive character of the {pair} depletion potential. For instance, while the {coarse-grained} approximation predicts a monotonic decrease of $B_3^{(1)}$ with increasing $q$, the exact coefficient presents a non-monotonic behavior with a minimum at $q\simeq 0.54$. Also, $B_4^{(1)}$ is negative definite in the {coarse-grained} approximation, while it actually changes from negative to positive at $q\simeq 0.74$. {As for the FV theory, it qualitatively agrees with the main trends of the exact coefficients $B_3^{(1)}$ and $B_4^{(1)}$, especially as $q$ increases.}

\begin{figure}
\includegraphics[width=8cm]{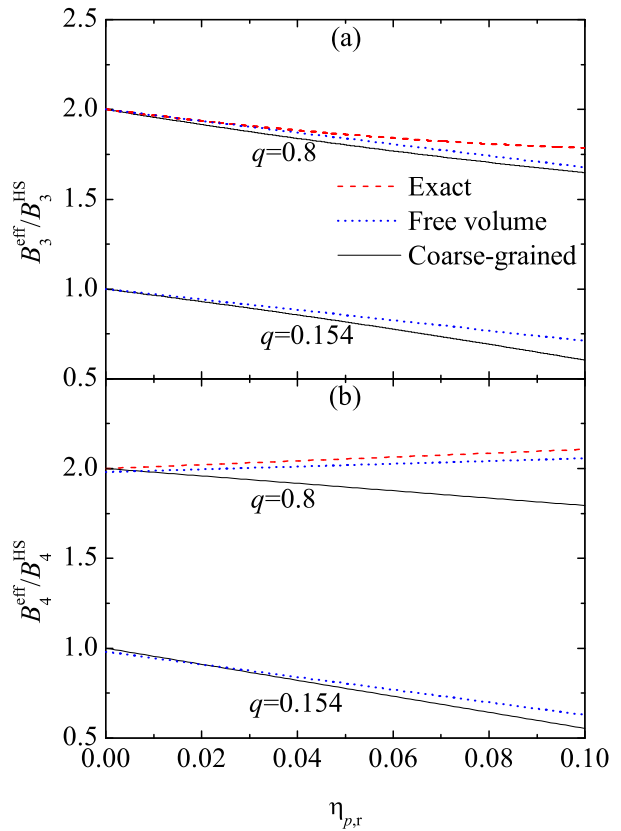}
\caption{Plot of (a) $B_3^\eff(\etar)\approx B_3^\hs+B_3^{(1)}\etar+B_3^{(2)}\etar^2$ and (b) $B_4^\eff(\etar) \approx B_4^\hs+B_4^{(1)}\etar$ for $q=0.154$ and $q=0.8$. To aid visibility, the curves for $q = 0.8$ have been shifted vertically by $1.0$.}
\label{fig2}
\end{figure}

Assuming sufficiently small values of $\etar$,  expansion \eqref{2} can be truncated to obtain the approximate forms $B_3^\eff(\etar) \approx B_3^\hs+B_3^{(1)}\etar+B_3^{(2)}\etar^2$ and $B_4^\eff(\etar) \approx B_4^\hs+B_4^{(1)}\etar$. The resulting curves for $q=0.154\lesssim q_0$ and $q=0.8$ in the range $0\leq\etar\leq 0.1$ are plotted in Fig.\ \ref{fig2}. Figure \ref{fig2}(a) is qualitatively analogous to Fig.\ 3 of Ref.\ \onlinecite{AW14b}. We observe that {the impact of} three- and four-body interactions {on $B_3^\eff$ and $B_4^\eff$} becomes relevant for $\etar\gtrsim {0.02}$ if $q=0.8$.

\begin{figure}
\includegraphics[width=8cm]{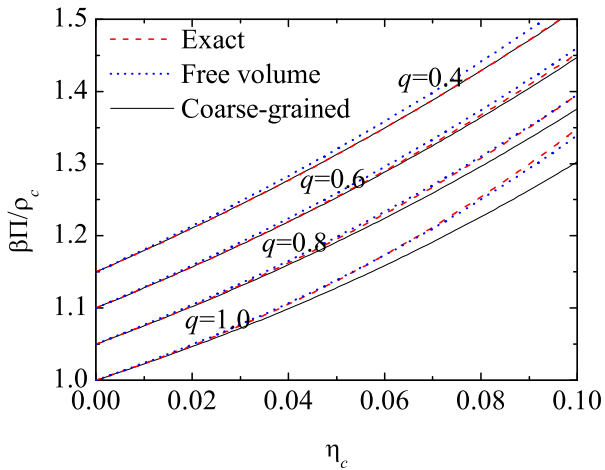}
\caption{{Plot of the osmotic compressibility factor $\beta\Pi/\rho_c$ versus the colloid packing fraction $\eta_c$ for $\etar=0.1$ and $q=0.4$, $0.6$, $0.8$, and $1.0$. To aid visibility, the curves for $q = 0.4$, $0.6$, and $0.8$ have been shifted vertically by $0.05$, $0.10$, and $0.15$, respectively.}}
\label{fig3}
\end{figure}

{Notwithstanding the results displayed in Figs.\ \ref{fig1} and \ref{fig2}, and since the coarse-grained description gives the exact effective second virial coefficient $B_2^\eff$,  the osmotic pressure of the colloids in the presence of nonadsorbing polymers is expected to be well described by the coarse graining method if the behavior is dominated by the second virial coefficient. To clarify this point, we consider the expansions \eqref{osmotic} and \eqref{2} truncated for $n+j\geq 6$, which implies small values of both packing fractions $\eta_c$ and $\etar$. Figure \ref{fig3} shows the corresponding compressibility factor $\beta\Pi/\rho_c$ within the range $0\leq\eta_c\leq 0.1$ for $\etar=0.1$ and several values of the size ratio $q$. Under those conditions, no difference between the exact and coarse-grained results are visible for $q=0.4$ and very small deviations can be observed for $q=0.6$ near $\eta_c=0.1$. Only for high size ratios ($q=0.8$ and $1.0$) it is apparent that the coarse-grained approximation underestimates the osmotic pressure; an effect that is expected to become more and more important as the packing fractions $\eta_c$ and $\etar$ increase beyond the range of applicability of the truncation for $n+j\geq 6$. It is interesting to note that the FV theory is rather close to the exact results for $q=0.8$ and $1.0$.}

\section{Concluding Remarks}
\label{sec6}

Using the available results\cite{HTSYFS15} for the virial coefficients of the AO binary-mixture model,  we have assessed, for size ratios $0<q<1$, the effect of neglecting three- and four-body interactions on the values of the effective one-component virial coefficients $B_3^\eff$ and $B_4^\eff$  that follow from the depletion pair potential derived in the coarse-grained version of such a model. While it was already well known that the coarse-grained version is exact for $q\leq q_0= 2/\sqrt{3}-1\simeq 0.1547$, the mapping between the virial coefficients of the true mixture and the effective ones that we have presented here, together with the corresponding analytical results, have allowed us to explicitly quantify the differences for the partial contributions $B_3^{(1)}$, $B_3^{(2)}$, and $B_4^{(1)}$ for $q>q_0$.
As an extra bonus of this mapping,  exact analytical expressions for the binary-mixture coefficients $E_{11122}$ and $E_{11112}$ were derived for any size ratio $q<q_0$. The same was in turn useful to check the accuracy of our numerical results  for those coefficients, which were proven to be very reliable.

The results indicate that the {coarse-grained} pair depletion approximation is very accurate for $q_0<q\lesssim 0.4$ but  one must certainly take into account the influence of three-body interactions on $B_3^\eff$ and $B_4^\eff$ if $q> 0.6$,  their role becoming essential as $q$ increases. While it is not possible at this stage to disentangle the roles of three- and four-body interactions on $B_4^\eff$, it is reasonable to expect that four-body terms could be important at least for values of $q$ close to unity.  All these facts should be especially noteworthy when dealing with dense systems. Also, for small values of $\etar$, such an influence has been shown here to be relevant. In fact, as already pointed out in the case of three-body interactions by the numerical studies of Ashton and Wilding,\cite{AW14a,AW14b} the deviation between the exact and {coarse-grained} values of  $B_3^\eff$  significantly increases as $\etar$ becomes larger. Therefore, care must be exercised when drawing conclusions from the coarse-grained version of the AO model if either $q$ or $\etar$, or both, are large.

\begin{acknowledgments}
The research of A.S. has been partially supported by the Spanish Government through Grant No.\ FIS2013-42840-P and by the Regional Government of Extremadura (Spain) through Grant No.\ GR15104 (partially financed by ERDF funds).
\end{acknowledgments}

\appendix

\section{{The FV theory}}
\label{appFV}
{In the FV theory\cite{LPPSW92,BVS14} the free energy of the system is expressed as a sum of a term corresponding to a pure colloidal suspension in the volume $V$  and a term corresponding to a pure polymer solution in the volume $\alpha(\eta_c)V$, where the free volume fraction $\alpha(\eta_c)$ is motivated by scaled particle theory. The corresponding equation of state is\cite{LPPSW92}}
 \bal
{\frac{\beta p}{\rho}}=&{x_c Z_{\text{CS}}(\eta_c)+\frac{x_p}{1-\eta_c}+\frac{x_p q \eta_c}{(1-\eta_c)^2}\Big[3+3q+q^2}
 \nn
   &{+3q(3+2q)\frac{\eta_c}{1-\eta_c}+9q^2\frac{\eta_c^2}{(1-\eta_c)^2}\Big],}
   \label{FV}
  \eal
{where  $Z_{\text{CS}}(\eta)=(1+\eta+\eta^2-\eta^3)/(1-\eta)^3$ is the Carnahan--Starling compressibility factor of a {one-component} HS fluid. Equation \eqref{FV}  is consistent with the exact second and third virial coefficients [see Eqs.\ \eqref{B2}, \eqref{B3}, \eqref{B11}, and \eqref{C112}]. On the other hand, the FV fourth and fifth virial coefficients are approximate only. They are given by Eqs.\ \eqref{B4} and \eqref{B5} with}
\beq
{D_{1111}=\left(\frac{\pi}{6}\right)^3 18,\quad E_{11111}=\left(\frac{\pi}{6}\right)^4 28,}
\label{D1111FV}
\eeq
\beq
{D_{1112}(q)=\left(\frac{\pi}{6}\right)^3\frac{1+9q+36q^2+30q^3}{4},}
\label{D1112FV}
\eeq
\beq
{E_{11112}(q)=\left(\frac{\pi}{6}\right)^4\frac{1+12q+66q^2+76q^3}{5},}
\label{E11112FV}
\eeq
\beq
{D_{1122}(q)=E_{11122}(q)=E_{11222}(q)=0.}
\label{D1122FV}
\eeq
{Insertion into Eqs.\ \eqref{8}--\eqref{14} yields}
\bal
{B_3^{(1)}(q)=}&{-\left(\frac{\pi}{6}\right)^2\Big(24+33q-3q^2-20q^3-12q^4}\nn
&{-3q^5-\frac{q^6}{3}\Big),}
\label{B31FV}
\eal
\bal
{B_4^{(1)}(q)=}&{-\left(\frac{\pi}{6}\right)^3\Big(54+\frac{333q}{4}-72q^2-147q^3-63q^4}\nn
&{+\frac{117q^5}{8}+\frac{43q^6}{2}+\frac{33q^7}{4}+\frac{3q^8}{2}+\frac{q^9}{8}\Big),}
\label{B41FV}
\eal
\beq
{B_2^{(2)}=B_2^{(3)}=B_3^{(2)}=0.}
\label{B22FV}
\eeq
{Thus, only the coefficients $B_2^{(0)}$, $B_2^{(1)}$, and $B_3^{(0)}$ are exactly given by the FV theory.}

\section{Derivation of Eqs.\ \protect\eqref{6}--\protect\eqref{14}}
\label{appA}
{We start by rewriting Eq.\ \eqref{comp} to fifth order as}
  \bal
 {\beta} p(\rho_c,\rho_p)=&\rho_c+\rho_p+B_{11}\rho_c^2+2B_{12}\rho_c\rho_p+C_{111}\rho_c^3\nn
  &+3C_{112}\rho_c^2\rho_p+D_{1111}\rho_c^4+4D_{1112}\rho_c^3\rho_p\nn
  &+6D_{1122}\rho_c^2\rho_p^2+E_{11111}\rho_c^5+5E_{11112}\rho_c^4\rho_p\nn
  &+10E_{11122}\rho_c^3\rho_p^2{+10E_{11222}\rho_c^2\rho_p^3}+\mathcal{O}(\rho^6),
    \label{3}
  \eal
where we have used Eqs.\ \eqref{B2}--\eqref{B5}. Next, from Eq.\ \eqref{free_e} and the thermodynamic relation $\mu_p=(\partial a/\partial \rho_p)_{\rho_c}$, we obtain
\beqa
{\beta}\mu_p&=&\ln\left(\rho_p{\Lambda_p^3}\right)+2B_{12}\rho_c+\frac{3}{2}C_{112}\rho_c^2+\frac{4}{3}D_{1112}\rho_c^3\nn
&&+4D_{1122}\rho_c^2\rho_p
+\frac{5}{4}E_{11112}\rho_c^4+5E_{11122}\rho_c^3\rho_p\nn
&&{+\frac{15}{2}E_{11222}\rho_c^2\rho_p^2}+\mathcal{O}(\rho^5).
  \label{4}
\eeqa
Consequently,    fugacity {\eqref{zp}} can be written as
\bal
\frac{z_p}{\rho_p}=&1+2B_{12}\rho_c+\left(2B_{12}^2+\frac{3}{2}C_{112}\right)\rho_c^2
+\left(\frac{4}{3}B_{12}^3\right.\nn
&
\left.+3B_{12}C_{112}+\frac{4}{3}D_{1112}\right)\rho_c^3+4D_{1122}\rho_c^2\rho_p\nn
&+\left(\frac{2}{3}B_{12}^4+3B_{12}^2C_{112}+\frac{8}{3}B_{12}D_{1112}+\frac{9}{8}C_{112}^2\right.\nn
&\left.+\frac{5}{4}E_{11112}\right)\rho_c^4
+\left(8B_{12}D_{1122}+5E_{11122}\right)\rho_c^3\rho_p
\nn&{+\frac{15}{2}E_{11222}\rho_c^2\rho_p^2}
+\mathcal{O}(\rho^5).
  \label{5}
\eal
This can be inverted to express $\rho_p$ as a series expansion in powers of $\rho_c$ and $z_p$,
\bal
\frac{\rho_p}{z_p}=&1-2B_{12}\rho_c+\left(2B_{12}^2-\frac{3}{2}C_{112}\right)\rho_c^2
-\left(\frac{4}{3}B_{12}^3\right.\nn
&
\left.-3B_{12}C_{112}+\frac{4}{3}D_{1112}\right)\rho_c^3-4D_{1122}\rho_c^2z_p\nn
&+\left(\frac{2}{3}B_{12}^4-3B_{12}^2C_{112}+\frac{8}{3}B_{12}D_{1112}+\frac{9}{8}C_{112}^2\right.\nn
&\left.-\frac{5}{4}E_{11112}\right)\rho_c^4
+\left(16B_{12}D_{1122}-5E_{11122}\right)\rho_c^3z_p\nn
&{-\frac{15}{2}E_{11222}\rho_c^2z_p^2}+\cdots,
  \label{5b}
\eal
where the ellipsis denotes terms of order $\rho_c^nz_p^j$ with $n+j\geq 6$.

Inserting Eqs.\ \eqref{3} and \eqref{5b} into Eq.\ \eqref{1} one can easily identify the coefficients shown in Eqs.\ \protect\eqref{6}--\protect\eqref{14}.

\begin{widetext}
\section{Expressions for $B_4^{(1)}$ in the {coarse-grained} approximation}
\label{appB}
In the range of interest $1<r<1+q<2$, the expressions for the contributions of $y_2^\hs(r)$ in Eq.\ \eqref{y2HS} are\cite{NvH52,SM07}
\beq
\varphi(r)\equiv\DDE=-\frac{\pi^2}{36}\frac{(r-3)^4}{35r}(r^3+12r^2+27r-6),
\label{n7}
\eeq
\beq
\psi(r)\equiv\DDG=\frac{\pi^2}{36}\frac{(r-2)^2}{35r}\left(r^5+4r^4-51r^3-10r^2+479r-81\right),
\label{n10}
\eeq
\bal
\chi(r)\equiv\DDI+\DDL=&\pi\Theta\left(\sqrt{3}-r\right)\Bigg[-r^2\left(\frac{3r^2}{280}-\frac{41}{420}\right)\sqrt{3-r^2}
-\left(\frac{23}{15}r-\frac{36}{35r}\right)
\cos^{-1}\frac{r}{\sqrt{3(4-r^2)}}
\nn
& +\left(\frac{3
r^6}{560}-\frac{r^4}{15}+\frac{r^2}{2}+\frac{2r}{15}-\frac{9}{35r}\right)\cos^{-1}\frac{r^2+r-3}{\sqrt{3(4-r^2)}}
\nn &
+\left(\frac{3
r^6}{560}-\frac{r^4}{15}
+\frac{r^2}{2}-\frac{2r}{15}+\frac{9}{35r}\right)
\cos^{-1}\frac{-r^2+r+3}{\sqrt{3(4-r^2)}}\Bigg].
\label{n20}
\eal
Use of Eqs.\ \eqref{n7}--\eqref{n20} into Eq.\ \eqref{B41} gives
\beq
B_4^{(1)}=B_{4,\varphi\psi}^{(1)}+B_{4,\chi}^{(1)},
\label{B41depl}
\eeq
where
\beq
B_{4,\varphi\psi}^{(1)}=-\frac{\pi^3}{3}\left(1+\frac{q}{8}-q^2-\frac{7q^3}{24}+\frac{29q^4}{140}+\frac{41q^5}{560}-
\frac{79q^6}{5040}-\frac{37q^7}{16\,800}
+\frac{q^8}{2100}+\frac{q^9}{25\,200}\right),
\label{B4phipsi}
\eeq
\beq
B_{4,\chi}^{(1)}=B_{4,\chi A}^{(1)}+\Theta\left(\sqrt{3}-1-q\right)\Delta B_{4,\chi}^{(1)},
\label{B41total}
\eeq
with
\bal
B_{4,\chi A}^{(1)}=&-\frac{\pi ^2}{22\,400 q^3} \Bigg[\frac{2\sqrt{2}}{105}\left(375\,323+1\,710\,828 q+548\,814 q^2-204\,400 q^3\right)-{\pi}   \left(23\,669+76\,404
   q\right.\nn
   &\left.+56\,562 q^2+12\,240 q^3\right)+{135} \left(387+1164 q+990 q^2+272 q^3\right) \cos
   ^{-1}\frac{1}{3}\Bigg],
   \label{B4A}
   \eal
\bal
\Delta B_{4,\chi}^{(1)}=&
\frac{\pi ^2}{11\,200 q^3} \Bigg[\frac{\sqrt{2-2q-q^2}}{105}
    \left(375\,323+1\,599\,922 q+413\,153 q^2-465\,800 q^3-241\,510 q^4-108\,524
   q^5\right.\nn
   &\left.-40\,954 q^6-4760 q^7+2555 q^8+1050 q^9+105 q^{10}\right)
   +\frac{(1-q)^5}{2} \left(1603 + 2243 q + 1479 q^2 + 855 q^3  \right.\nn
   &\left.+ 405 q^4 + 117 q^5+ 17 q^6 + q^7\right) \cos^{-1}\frac{2 q-1}{\sqrt{3} (1-q)}
   -8 (1+q)^3\left(1373 - 336 q - 72 q^2 + 96 q^3 + 24 q^4\right)\nn
   &\times  \cos
   ^{-1}\frac{1+q}{\sqrt{3}}
   +\frac{(3+q)^5}{2} \left(111 + 171 q - 117 q^2 + 135 q^3 + 25 q^4 - 3 q^5 - 3 q^6 + q^7
   \right) \cos^{-1}\frac{5+2 q}{\sqrt{3} (3+q)}\Bigg].
   \label{DeltaB4b}
\eal

Taking into account that \eqref{B41total} is exact for $q<q_0$, and using Eqs.\ \eqref{B11}, \eqref{C112}, \eqref{6a}, and \eqref{14}, one can obtain the following exact expression of $E_{11112}$ for $q<q_0$:
   \bal
E_{11112}=&\left(\frac{\pi}{6}\right)^4 \frac{1}{5}\left(
1+12 q+66 q^2  +40
   q^3+81 q^4 +\frac{108 q^5}{5}-144 q^6-\frac{6516 q^7}{35}-\frac{1521 q^8}{10}-\frac{3384
   q^9}{35}-\frac{6858 q^{10}}{175}\right.\nn
  & \left.-\frac{1458 q^{11}}{175}-\frac{243 q^{12}}{350}\right)+\frac{2\pi q^3}{45}\left(B_{4,\chi A}^{(1)}+\Delta B_{4,\chi}^{(1)}\right).
\label{E11112}
   \eal
   \end{widetext}

\end{document}